\documentclass{aastex631}

\usepackage{amsmath}
\usepackage[title]{appendix}

\usepackage{xcolor, microtype}
\definecolor{twitterblue}{RGB}{64,153,255}
\definecolor{linkcolor}{rgb}{0.1216,0.4667,0.7059}

\usepackage{ mathrsfs }
\usepackage[T1]{fontenc}
\usepackage{hyperref}
\usepackage{comment}
\let\tablenum\relax
\usepackage{siunitx}
\newcommand{\ktwo}{\emph{K2}}

\newcommand{\rearth}{$R_{\oplus}$}

\newcommand{\rsun}{$R_{\odot}$}
\newcommand{\msun}{$M_{\odot}$}

\begin{document}

\title{Scaling K2 VII: Evidence for a high occurrence rate of hot sub-Neptunes at intermediate ages}

\correspondingauthor{Jessie L. Christiansen}
\email{christia@ipac.caltech.edu}

\author[0000-0002-8035-4778]{Jessie L. Christiansen}
\affiliation{NASA Exoplanet Science Institute, IPAC, MS 100-22, Caltech, 1200 E. California Blvd, Pasadena, CA 91125}

\author[0000-0003-1848-2063]{Jon K. Zink}\thanks{NASA Hubble Fellow}
\affiliation{Department of Astronomy, Caltech, 1200 E. California Blvd, Pasadena, CA 91125}

\author[0000-0003-3702-0382]{Kevin K. Hardegree-Ullman}
\affiliation{Steward Observatory, The University of Arizona, Tucson, AZ 85721, USA}

\author[0000-0002-3853-7327]{Rachel B. Fernandes}
\altaffiliation{President's Postdoctoral Fellow}
\affil{Department of Astronomy \& Astrophysics, Center for Exoplanets and Habitable Worlds, The Pennsylvania State University, University Park, PA 16802, USA}
\affil{Lunar and Planetary Laboratory, The University of Arizona, Tucson, AZ 85721, USA}

\author[0000-0003-3729-1684]{Philip F. Hopkins}
\affiliation{TAPIR, MS 350-17, Caltech, 1200 E. California Blvd, Pasadena, CA 91125}

\author[0000-0003-3729-1684]{Luisa M. Rebull}
\affiliation{Infrared Science Archive (IRSA), IPAC, MS 100-22, Caltech, 1200 E. California Blvd, Pasadena, CA 91125, USA}

\author[0000-0001-8153-639X]{Kiersten M. Boley}
\altaffiliation{NSF Graduate Research Fellow}
\affiliation{Department of Astronomy, The Ohio State University, Columbus, OH 43210, USA}

\author[0000-0003-4500-8850]{Galen J. Bergsten}
\affil{Lunar and Planetary Laboratory, The University of Arizona, Tucson, AZ 85721, USA}

\author[0000-0001-6381-515X]{Sakhee Bhure}
\altaffiliation{Volunteer Researcher}
\affiliation{NASA Exoplanet Science Institute, IPAC, Caltech, 1200 E. California Blvd, Pasadena, CA 91125}



\begin{abstract}

The NASA \ktwo\ mission obtained high precision time-series photometry for four young clusters, including the near-twin 600--800~Myr-old Praesepe and Hyades clusters. Hot sub-Neptunes are highly prone to mass-loss mechanisms, given their proximity to the the host star and the weakly bound gaseous envelopes, and analyzing this population at young ages can provide strong constraints on planetary evolution models. Using our automated transit detection pipeline, we recover 15 planet candidates across the two clusters, including 10 previously confirmed planets. We find a hot sub-Neptune occurrence rate of 79--107\% for GKM stars in the Praesepe cluster. This is 2.5--3.5$\sigma$ higher than the occurrence rate of $16.54_{-0.98}^{+1.00}$\% for the same planets orbiting the $\sim$3--9~Gyr-old GKM field stars observed by \ktwo, even after accounting for the slightly super-solar metallicity ([Fe/H]$\sim$0.2 dex) of the Praesepe cluster. We examine the effect of adding $\sim$100 targets from the Hyades cluster, and extending the planet parameter space under examination, and find similarly high occurrence rates in both cases. The high occurrence rate of young, hot sub-Neptunes could indicate either that these planets are undergoing atmospheric evolution as they age, or that planetary systems that formed when the Galaxy was much younger are substantially different than from today. Under the assumption of the atmospheric mass-loss scenario, a significantly higher occurrence rate of these planets at the intermediate ages of Praesepe and Hyades appears more consistent with the core-powered mass loss scenario sculpting the hot sub-Neptune population, compared to the photoevaporation scenario.

\end{abstract}

\keywords{Exoplanet evolution (491) --- Transit photometry (1709) --- Hot Neptunes (754) --- Interdisciplinary astronomy(804)}


\section{Introduction} \label{sec:intro}

After successive hardware failures rendered the NASA \emph{Kepler} spacecraft \citep{Borucki2010} incapable of continuing to point with high precision and stability at its original field of view, the \emph{K2} mission repurposed the spacecraft to perform a four year survey of 18 fields, called `campaigns', around the ecliptic plane \citep{Howell14}. This significant expansion in sky coverage compared to the original \emph{Kepler} mission led to a corresponding increase in both the number and variety of targets for which high precision time-series photometry could be obtained. Where the goal of \emph{Kepler} was to examine the planet occurrence rates for solar-type stars, the target list for \ktwo\ was provided by the community to meet a variety of science goals. The goal of the Scaling K2 project is the exploitation of this expanded target set to search for planet candidates in new areas of stellar parameter space beyond that probed by \emph{Kepler}. As the \ktwo\ campaigns were limited to 70--80 days in duration, and the photometric noise achieved with \ktwo\ was higher than that achieved with \emph{Kepler}, our sensitivity is necessarily limited to short-period ($<$40 day) planets down to super-Earth-size. 

In Papers I--V we presented the uniformly derived set of stellar parameters for the \ktwo\ targets comprising our search sample \citep{har20}, the automated detection pipeline for transiting candidates, with well-characterized completeness and reliability \citep{zin20a}, a pilot study on a single campaign \citep{zin20b}, the full, homogeneous planet candidate catalog from campaigns 1--8 and 10--18 \citep{Zink2021}, and 60 newly validated planets from the catalog \citep{Christiansen2022}. In Paper VI \citep{Zink2023} we compared the FGK occurrence rates for \emph{Kepler} and \ktwo\ in their regions of stellar and planet parameter overlap and found good agreement between the two samples, validating our approach to date, and providing independent confirmation of the trends observed in \emph{Kepler} data. In addition, we found that the occurrence rate of small (sub-Neptunes and super-Earths), short (1--40 d) period planets is inversely correlated with the galactic oscillation amplitude of the host star---the higher a star reaches above the plane in its galactic orbit, the less likely it is to host small, short-period planets.

Another axis along which \ktwo\ can probe is that of stellar age---in its four-year mission it observed four young clusters: Upper Scorpius (5--10 Myr; $\sim$1100 members observed), Pleiades (120 Myr; $\sim$850 members), Hyades (600--800 Myr; $\sim$300 members) and Praesepe (600--800 Myr; $\sim$1000 members). As different planet formation and evolution scenarios are associated with different timescales, measuring occurrence rates of planets as a function of age provides particularly useful constraints on the viability of these scenarios. One particular current area of interest is the origin of the planet radius valley and sub-Neptune desert for small, short-period planets \citep{ful17,VanEylen2018,har20}. The leading candidates for the explanation of this feature are planetary atmosphere evolution in the form of either photoevaporation \citep{Lopez2012,Lopez13,Owen2013,Owen2017, Rogers2021} or core-powered mass loss \citep{Ginzburg2016,gup19,Gupta2020,Gupta2021}. These mechanisms are expected to act over very different timescales; \cite{Ginzburg2016} suggested that analyzing occurrence rates of young planetary systems would be a useful way to discriminate between the two. In this paper, we measure the occurrence rate for small, short-period planets (hot sub-Neptunes) in the Praesepe cluster, subsequently augmenting the analysis with the significantly smaller number of Hyades cluster members. The paper is organized as follows: in Section \ref{sec:candidates} we describe the cluster stellar sample and corresponding planet candidates retrieved by our pipeline; in Section \ref{sec:occrates} we describe the occurrence rate calculations; and in Section \ref{sec:disc} we discuss the implications of the results.

\begin{figure}[t!]
\includegraphics[width=0.495\textwidth]{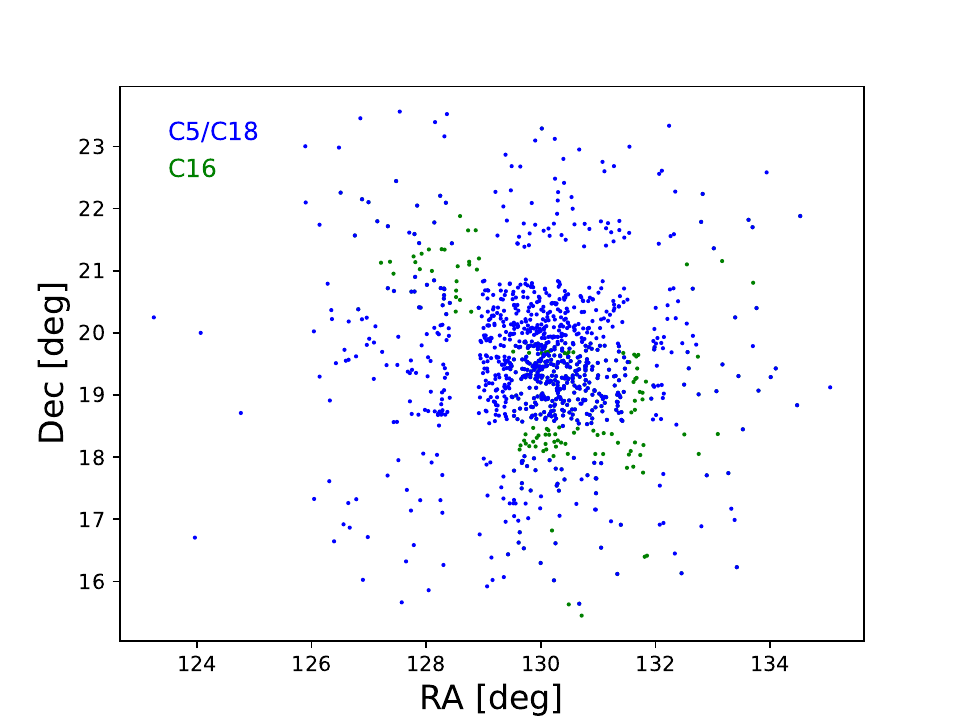}
\includegraphics[width=0.495\textwidth]{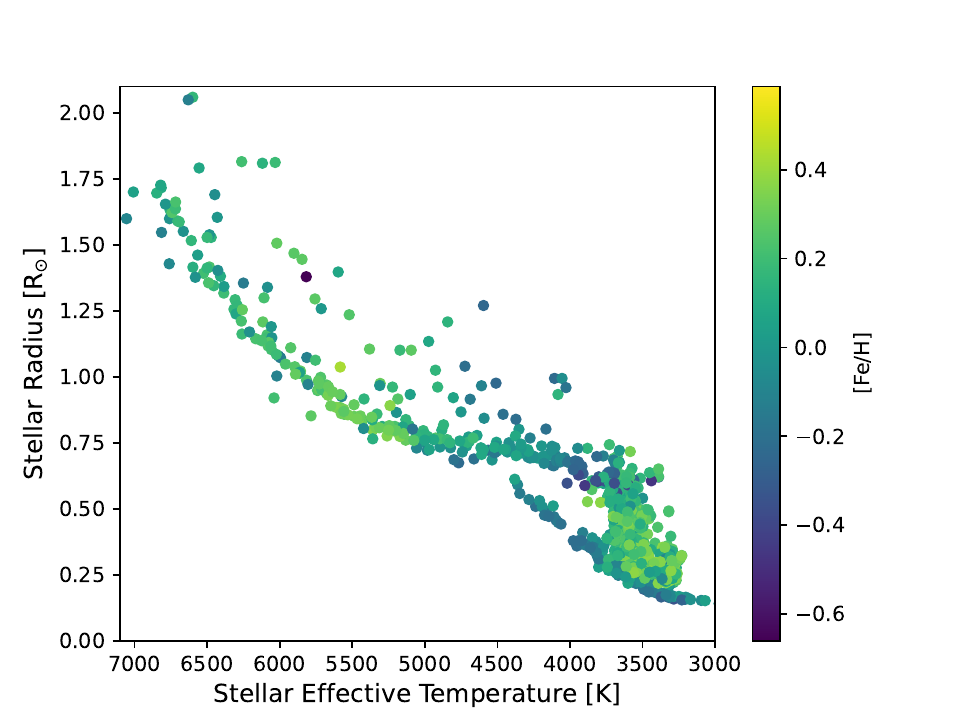}
\caption{\emph{Left:} The spatial distribution of the Praesepe cluster members on the sky observed by \ktwo; targets observed in campaigns 5 and 18 are shown in blue, and targets observed in campaign 16 are shown in green. \emph{Right:} The temperature, radii, and metallicity of the Praesepe cluster members with 8-hour rms CDPP $<$ 1200~ppm. The stellar parameters are from the uniform catalog of \ktwo\ stellar parameters published in \cite{har20}, and were derived using a random forest regression on photometry, trained on a large set of targets with LAMOST spectra; where the machine learning parameters were unavailable, they were supplemented with parameters from isochrone fitting. \label{fig:praesepe_locations}}
\end{figure}

\section{Sample definition} \label{sec:candidates}

\subsection{Praesepe}

Praesepe (NGC 2632, M44) is a 600--800~Myr-old open cluster located about 180 pc away. The age, distance, and metallicity of the cluster have been studied extensively \cite[see, e.g.,][and references therein]{Brandt2015,Gossage2018}. The cluster has slightly super-solar metallicity, with estimates ranging from [Fe/H] = 0.12$\pm$0.04 \citep{Boes2013} to 0.21$\pm$0.01 \citep{Dorazi2020}, depending on the calibration and methods used. Fortuitously lying in the ecliptic, it was observed by \ktwo\ in campaigns 5, 16, and 18 (the latter of which had the same pointing as campaign 5). We use the member list compiled by \cite{Rebull2017} in their analysis of the rotation periods of Praesepe members in \ktwo\ data, which originally included the targets observed in campaign 5 (and 18), and was subsequently augmented by additional members observed in campaign 16. Figure \ref{fig:praesepe_locations} shows the spatial distribution of the 1030 Praesepe targets observed by \ktwo\ across the sky.

\begin{figure}[t!]
\includegraphics[width=0.495\textwidth]{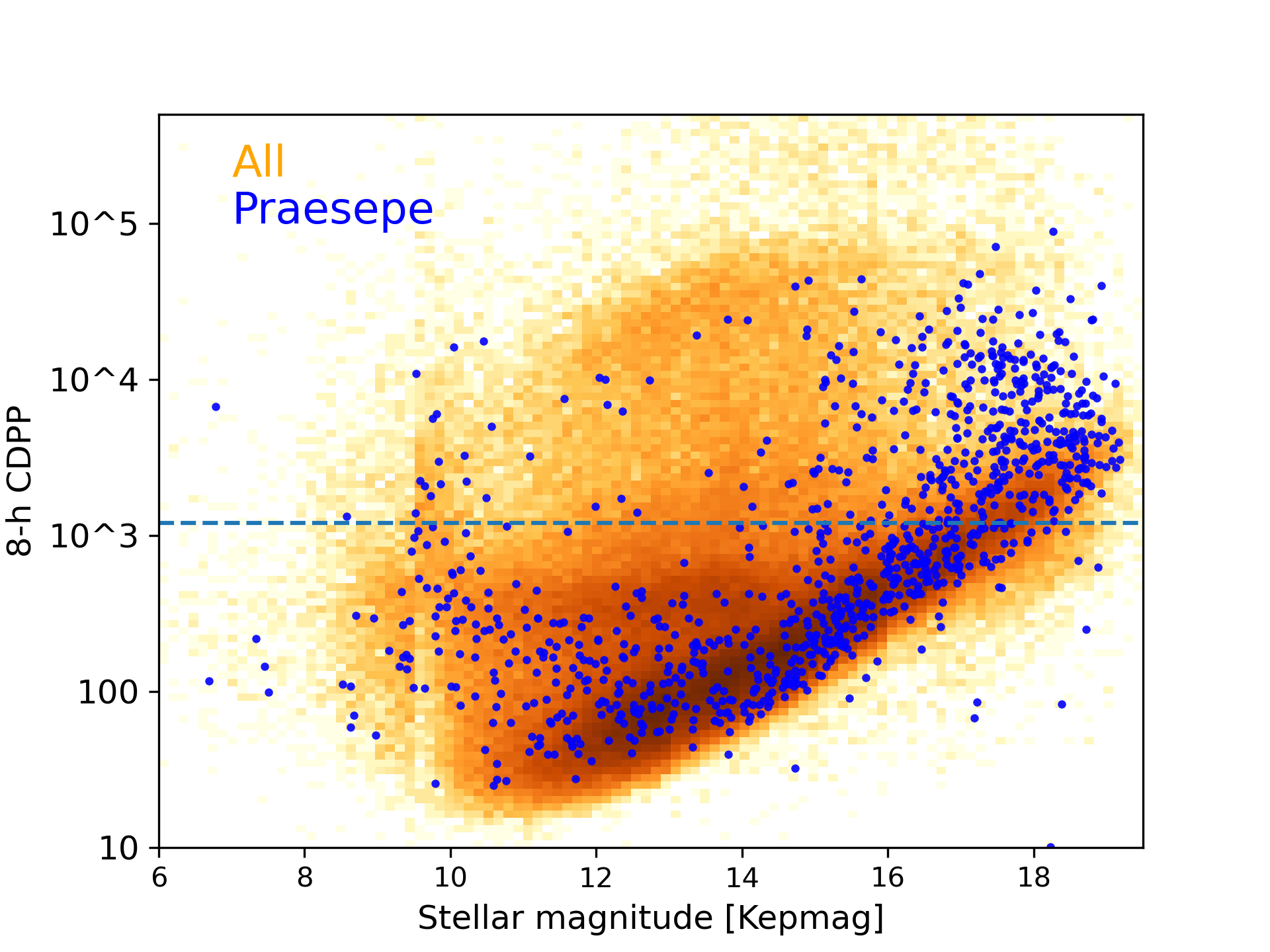}
\includegraphics[width=0.495\textwidth]{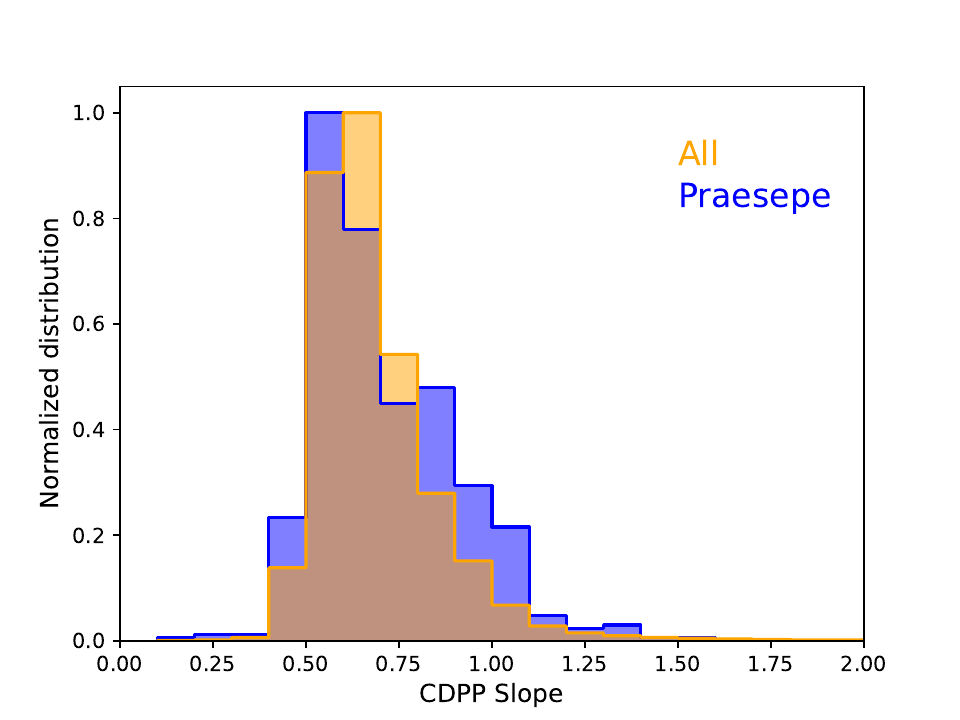}
\caption{\emph{Left:} The distribution of 8-hour rms CDPP values for the Praesepe targets in blue, and for the full \ktwo\ dataset in orange. The 1200-ppm cut-off is shown as the dashed line---it primarily removes faint, late-type M dwarfs. \emph{Right:} The distribution of 8-hour/2-hour CDPP slope values for the targets below the 1200-ppm cut-off, with Praesepe shown in blue and the full \ktwo\ dataset in orange. For white noise, the CDPP slope is expected to be 0.5---both distributions peak just above 0.5 and lie primarily between 0.5 and 1.0, indicating non-pathological light curves. The Hyades targets we investigated follow comparable noise patterns.
\label{fig:praesepe_noise}}
\end{figure}

As part of our search of the full \ktwo\ dataset, presented in \cite{Zink2021}, the Praesepe targets were processed through our standard \ktwo\ data reduction and transit detection pipeline detailed in \cite{zin20a}; in brief, light curves are extracted using \texttt{EVEREST} \citep{lug16,lug18}, detrended using Gaussian process regression and harmonic removal, searched for periodic transit signals using \texttt{TERRA} \citep{pet13b}, and the resulting signals uniformly vetted using the \texttt{EDI-Vetter} tool \citep{zin20a}. Although there is some overlap between the campaigns, for this analysis we treat each campaign separately, rather than combine light curves from multiple campaigns; this limits our period sensitivity to below 40 days but greatly simplifies the completeness and reliability corrections in the subsequent analysis. We compared the noise properties of the Praesepe targets to the field \ktwo\ field star dataset (after removing the known young stars in Upper Scorpius, Pleiades, Hyades and Praesepe)---young stars rotate more rapidly and therefore, although the Praesepe members studied here are largely on the main sequence, their light curves could potentially have increased correlated noise. This could make transit signals more difficult to detect than would be reflected by the detection efficiency previously measured for the pipeline across all targets, and potentially necessitate an updated detection efficiency measurement specifically for the Praesepe targets. We examine the Combined Differential Photometric Precision (CDPP) values \citep{Christiansen2012} established for \emph{Kepler} light curves. The left panel of Figure \ref{fig:praesepe_noise} shows the 8-hour root-mean-square (rms) CDPP values as a function of the \emph{Kepler} magnitude for the Praesepe targets in blue, overlaid on an orange heatmap showing the 8-hour rms CDPP values for our full \ktwo\ dataset. The Praesepe targets largely follow the broad features of the overall sample, except at the very faintest magnitudes. \cite{Zink2021} noted that stars with an 8-hour rms CDPP above 1200~ppm do not contribute meaningfully to the derived occurrence rates, as they have no sensitivity to the planet parameter range under study\footnote{To ensure that we are not changing the answer by removing these stars, we perform a test, denoted in Table \ref{tab:coefficients}, where we run the same analysis with and without the CDPP cut, and find that it does not significantly change the result}. Applying the same CDPP cut as \cite{Zink2021} removes 429 targets, largely very faint M dwarfs smaller than 0.33~\rsun. The 601 remaining Praesepe targets are shown in the right hand panel of Figure \ref{fig:praesepe_locations}.

In addition to the overall scatter in a light curve, we can also measure the extent of the \emph{correlated} noise. \cite{bur17} showed that calculating the CDPP ``slope,'' by measuring how the rms CDPP values change as a function of the timescale over which they are calculated, could capture this information. For white noise, the rms CDPP values should decrease with increasing duration as $\sqrt{N_{\rm short}/N_{\rm long}}$, where $N_{\rm short}$ and $N_{\rm long}$ are the shorter and longer durations over which the CDPP is calculated, respectively. The right panel of Figure \ref{fig:praesepe_noise} shows the normalized distributions of the CDPP slopes for the 2-hour and 8-hour rms CDPP values of the 601 Praesepe targets in blue and the remainder of the \ktwo\ targets in orange. White noise would produce a CDPP slope of $\sqrt{2/8}=0.5$, and both populations peak at just above that value (and drop off rapidly below it), indicating largely well-behaved light curves. The Praesepe stars have a slightly broader distribution up to a CDPP slope of $\sim$1, indicating a marginally higher proportion of light curves with correlated noise than the full dataset, although above $\sim$1.1 (where binning over a longer timescale actually \emph{increases} the noise, indicating highly correlated noise) the fractional proportion of the two populations is very similar. These comparisons indicate that the noise properties of the Praesepe targets are similar to the full \ktwo\ dataset and that the detection pipeline completeness and reliability measured on the full dataset in \cite{zin20b} are applicable here.

\begin{figure}
\begin{center}
\includegraphics[width=0.75\textwidth]{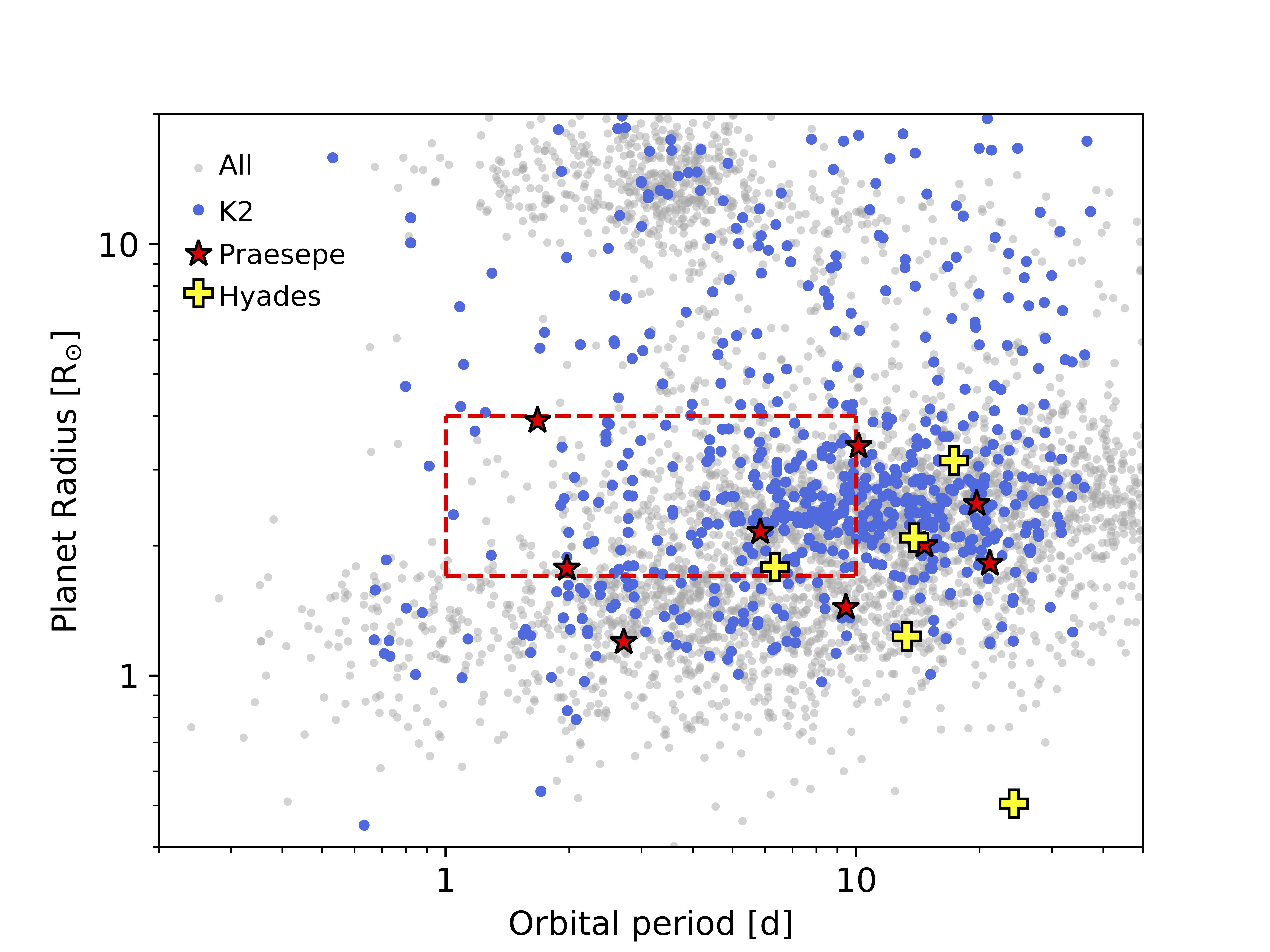}
\caption{Praesepe planet candidates are shown as red stars, and Hyades planet candidates are shown as yellow plus symbols. The full \ktwo\ planet candidate catalog of \cite{zin20a} is shown in blue, and all confirmed planets at the \cite{NEA} are shown in grey. The red box shows the hot sub-Neptune parameter space (1--10 days, 1.7--4~\rearth) analysed in Section \ref{sec:occrates}.\label{fig:praesepe_candidates}}
\end{center}
\end{figure}

The pipeline detected 10 planet candidates orbiting nine of the 601 Praesepe targets, originally published in the full \ktwo\ catalog provided in \cite{Zink2021}. The candidates are listed in Table \ref{tab:candidates}---they range from 1.2--3.9~\rearth, with periods from 1.67--21.2~days, typical of the large population of warm sub-Neptunes and super-Earths revealed by \emph{Kepler}. Given the interest in detecting and characterizing planets earlier in their formation and evolution histories, the \ktwo\ Praesepe data have been extensively searched for individual planets, and seven of the 10 candidates are already confirmed in the literature, as indicated in Table \ref{tab:candidates}. However, this paper presents the first statistical sample of planet candidates from Praesepe for population analysis. We note that one of the unconfirmed planet candidates (EPIC 211885995.01) has a consistency score falling below our cut-off of 0.75 \cite[see Section 3.13 of][]{zin20a}, and indeed a closer examination of the light curve indicates that the transit signal is likely related to stellar variability; we remove this target from further analysis, leaving nine remaining candidates shown in Figure \ref{fig:praesepe_candidates}. 

\subsection{Hyades}

In campaigns 4 and 13, \ktwo\ also observed the Hyades open cluster, a smaller, closer, near-twin of Praesepe. Age estimates range from 625$\pm$25 Myr \citep{Perryman1998}, to 680 Myr \citep{Gossage2018}, to 800 Myr \citep{David2015, Brandt2015}. It also has a slightly super-solar metallicity of [Fe/H] = 0.14$\pm$0.05 dex \citep{Perryman1998}, and has also been meticulously combed for planets. A list of potential Hyades members was assembled from membership lists in the literature \cite[e.g.][among others]{Douglas2016,Lodieu2019,Roser2011,Roser2019,Goldman2013}, the list of Hyades members submitted to the original \ktwo\ proposal call, and a run of \texttt{BANYAN} $\Sigma$ \citep{Gagne2018}, then compared to the list of objects with K2 light curves (Rebull et al. in prep), for a total of 300 targets. This is significantly smaller than our Praesepe target list, in part because Hyades is closer and therefore subtends a larger area on the sky, a smaller fraction of which falls in the \ktwo\ field of view; many Hyades cluster members were not observed by \ktwo. Applying the same noise and stellar parameter cuts as before leaves 101 Hyades targets to add to our sample. We examined the noise properties of these targets, as above for Praesepe, and found them to be comparable. Similarly to our Praesepe sample, the Hyades targets are on average cooler than the full \ktwo\ sample. 

Our pipeline recovered five planet candidates from these 101 Hyades targets, listed in Table \ref{tab:candidates}. Three are known planets, K2-155 b and c \citep{Hirano2018} and K2-136 c \citep{Mann2018}, although in both cases our pipeline missed known smaller or longer period additional known planets in the system. The final two are a pair of small (0.5~\rearth\ and 1.2~\rearth) candidates orbiting the faint ($m_{\rm Kep}=15.635$) target EPIC 246711015, which have thus far proven difficult to confirm. 


\begin{deluxetable*}{ccccccccc}
\tablecaption{Summary parameters of the 15 planet candidates and their host stars identified in Praesepe and Hyades. C\# = campaign in which the signal was detected by our pipeline. CP name = confirmed planet name in the literature. $^a$\cite{Mann2017}; $^b$\cite{Obermeier2016}; $^c$\cite{Rizzuto2018};  $^d$\cite{Hirano2018}; $^e$\cite{Mann2018}; $^*$removed from further analysis due to low consistency score.\label{tab:candidates}}
\tablehead{\colhead{EPIC ID} & \colhead{C\#} & \colhead{CP name} & \colhead{R$_{\rm p}$} & \colhead{Orbital Period} & \colhead{R$_{*}$} & \colhead{$T_{\rm eff}$} & \colhead{[Fe/H]} & \colhead{Sp.T}  \\
\colhead{} & \colhead{} & \colhead{} & \colhead{[R$_{\oplus}$]} & \colhead{[d]} & \colhead{[R$_{\odot}$]} & \colhead{[K]} & \colhead{[dex]} & }
\startdata
 \multicolumn{7}{c}{Praesepe} \\
211822797.01     & 16  & K2-103 b$^a$ & $1.819_{-0.090}^{+0.101}$ & $21.171661_{-0.003290}^{+0.003413}$ & $0.598\pm0.018$ & $3806\pm138$ & $0.110\pm0.235$ & M1\\
211885995.01     & 18  & $^*$ & $3.087_{-0.131}^{+0.142}$ & $9.863718_{-0.006848}^{+0.005667}$ & $0.605\pm0.019$ & $3676\pm138$ & $0.098\pm0.235$ & M1\\
211913977.01     & 16  & K2-101 b$^a$ & $2.006_{-0.091}^{+0.086}$ & $14.677689_{-0.001147}^{+0.001111}$ & $0.755\pm0.019$ & $4694\pm39$ & $0.109\pm0.036$ & K3\\
211916756.01     & 18  & K2-95 b$^b$ & $3.402_{-0.134}^{+0.134}$ & $10.135158_{-0.000635}^{+0.000698}$ &$0.415\pm0.014$ & $3576\pm138$ & $0.014\pm0.235$ & M3\\
211922849.01     &  5  & -- & $1.440_{-0.059}^{+0.066}$ & $9.443274_{-0.002157}^{+0.002052}$ & $0.517\pm0.016$ & $3652\pm138$ & $0.175\pm0.235$ & M2\\
211964830.01     & 16  & K2-264 b$^c$ & $2.153_{-0.105}^{+0.101}$ &  $5.839636_{-0.000527}^{+0.000477}$ & $0.473\pm0.015$ & $3594\pm138$ & $0.239\pm0.235$ & M3 \\
211964830.02     & 16  & K2-264 c$^c$ & $2.503_{-0.118}^{+0.117}$ & $19.662285_{-0.002070}^{+0.002742}$ & $0.473\pm0.015$ & $3594\pm138$ & $0.239\pm0.235$ & M3\\
211969807.01     &  5  & K2-104 b$^a$ & $1.773_{-0.078}^{+0.085}$ &  $1.974302_{-0.000192}^{+0.000168}$ & $0.505\pm0.017$ & $3693\pm138$ & $0.125\pm0.235$ & M2 \\
211990866.01     & 18  & K2-100 b$^a$ & $3.899_{-0.103}^{+0.101}$ &  $1.673847_{-0.000042}^{+0.000040}$ &  $1.208\pm0.020$ & $6116\pm16$ & $0.274\pm0.013$ & G1\\
212035441.01     & 18  & -- & $1.198_{-0.066}^{+0.063}$ &  $2.714743_{-0.000276}^{+0.000277}$ & $0.415\pm0.014$ & $3634\pm138$ & $0.310\pm0.235$ & M3 \\
\multicolumn{7}{c}{Hyades} \\
210897587.01 & 13 & K2-155 b$^d$ & $1.785_{-0.065}^{+0.066}$ & $6.34399_{-0.000050}^{+0.000044}$ & $0.584\pm0.019$ & $3893\pm56$ & $-0.831\pm0.051$ & K7 \\
210897587.02 & 13 & K2-155 c$^d$ & $2.088_{-0.075}^{+0.081}$ & $13.852825_{-0.000280}^{+0.000442}$ & $0.584\pm0.019$ & $3893\pm56$ & $-0.831\pm0.051$ & K7 \\
246711015.01 & 13 & -            & $1.233_{-0.094}^{+0.116}$ & $13.284219_{-0.001880}^{+0.002846}$ & $0.395\pm0.017$ & $3771\pm138$ & $-0.112\pm0.235$ & M1 \\
246711015.02 & 13 & -            & $0.505_{-0.084}^{+0.088}$ & $24.213118_{-0.003407}^{+0.003408}$ & $0.395\pm0.017$ & $3771\pm138$ & $-0.112\pm0.235$ & M1 \\
247589423.01 & 13 & K2-136 c$^e$ & $3.150_{-0.095}^{+0.099}$ & $17.306369_{-0.000152}^{+0.000171}$ & $0.725\pm0.018$ & $4231\pm41$ &  $-0.112\pm0.235$ & K5 \\
\enddata
\end{deluxetable*}

\section{Occurrence Rate Analysis} \label{sec:occrates}

\begin{figure}
\begin{center}
\includegraphics[width=0.49\textwidth]{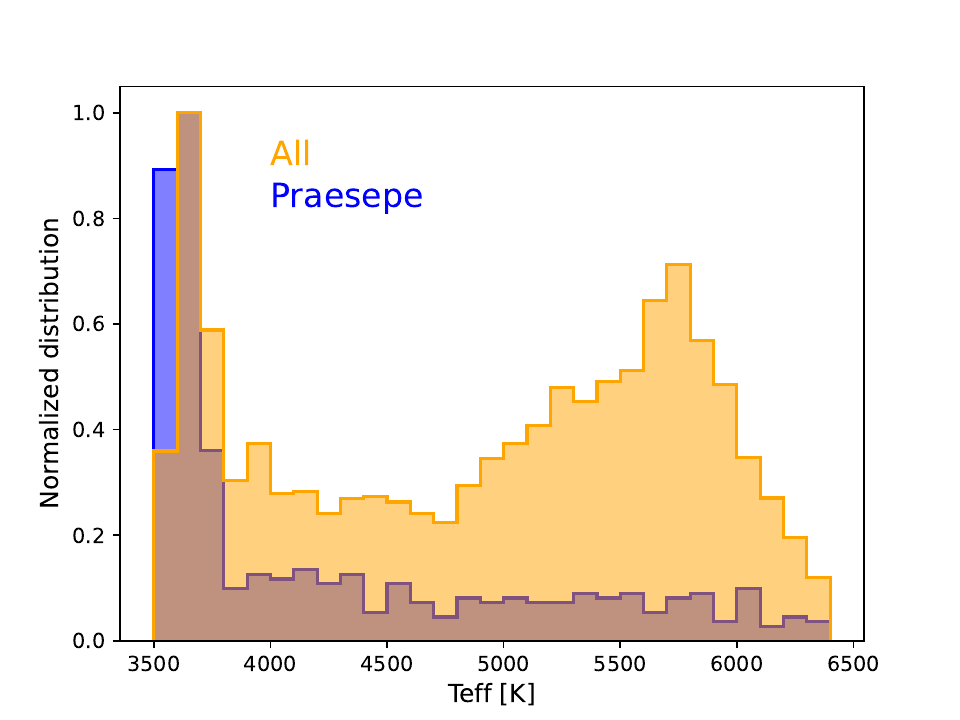}
\includegraphics[width=0.49\textwidth]{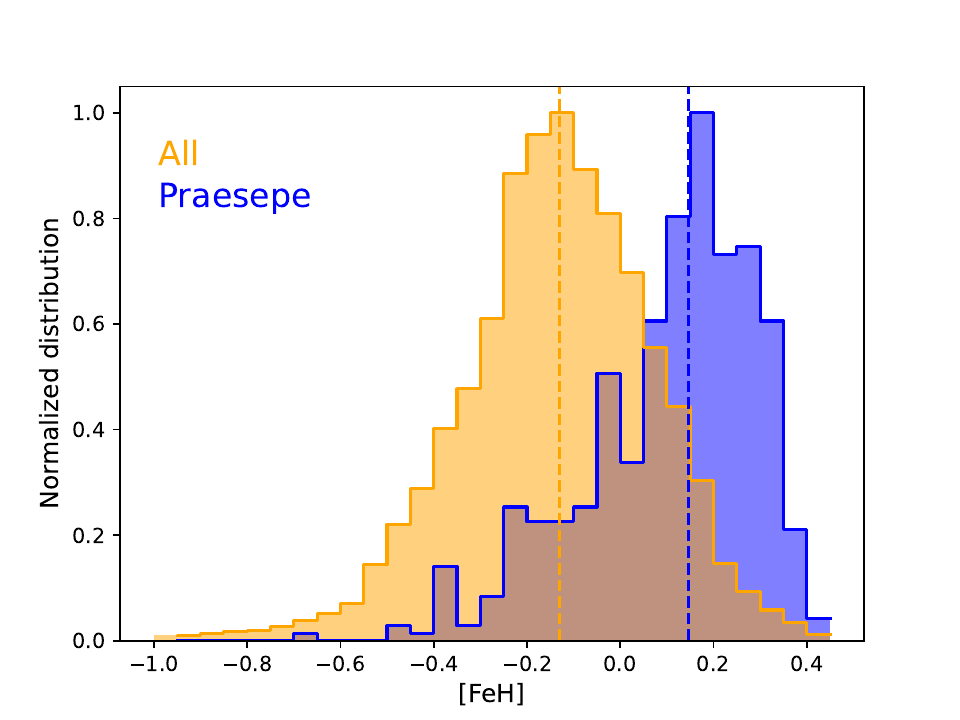}
\caption{The normalized distributions of the Praesepe GKM sample in orange and the field \ktwo\ GKM sample in blue, for stellar effective temperature (left) and stellar metallicity (right). The vertical dashed lines in the right panel are the median values of the distributions. The known super-solar metallicity of the Praesepe cluster is clear, although the distribution of our photometrically-derived values is wider than what is seen spectroscopically \citep{Boes2013,Dorazi2020}. The systematic differences in the stellar samples necessitate an appropriate parametric model that can account for any known correlations. \label{fig:praesepe_metallicity}}
\end{center}
\end{figure}

\subsection{Parametric Model}
\label{sec:model}

In order to calculate occurrence rates, we use the \texttt{ExoMult} forward modeling software \citep{zin19,zin20b}, which, in brief, produces synthetic \ktwo\ planet samples drawn from an underlying parametric planet distribution function, and simulates their detection throughput using the completeness measured in \cite{Zink2021}. The synthetic population of planets and the empirical sample were compared using the two-sample Andersen-Darling test statistic and then optimized for each relevant observable. Since the aim of the analysis is to compare the intermediate-age planet sample to the field \ktwo\ sample and to isolate any age correlation, careful attention must be paid to the form of the parametric model to ensure it captures the differences between the samples to the extent possible. 

Given its significantly larger dataset, we begin by examining the Praesepe sample. The Praesepe candidates orbit dwarf stars from G1--M3, so we restrict our population analysis to GKM stars spanning the range $0.33<R_{*}<1.5$~\rsun, $3500<T_{\rm eff}<6500$~K, $-0.5<{\rm [Fe/H]}<0.5$~dex, and log$g>$~4.0. Our stellar parameters were uniformly computed for the \ktwo\ sample as described in \cite{har20}, using a random forest regression on photometric colors trained on stars with LAMOST spectra, ultimately providing spectral types, effective temperatures, surface gravities, metallicities, radii, and masses. Where the machine learning parameters were unavailable, they were supplemented with parameters from isochrone grid matching, as per \cite{Zink2023}. For the well-behaved Praesepe targets, this reduces the dataset to nine planet candidates and 487 stars. For the well-behaved (8hr CDPP $<$1200 ppm) targets in the full \ktwo\ field star dataset (with the stars in the four identified young clusters removed), this leaves 520 planet candidates and 108,502 stars, refered to as the `field \ktwo\ sample`. The rest of the analysis will build on these two samples. 

The two stellar samples have significantly different distributions in stellar temperature and metallicity, shown in Figure \ref{fig:praesepe_metallicity}. There are well established correlations between the frequencies of planets over these stellar parameters. The median iron abundance of our Praesepe sample is 0.146 dex, consistent with the slightly super-solar metallicity of the cluster. Although sub-Neptunes do not have as steep a metallicity correlation as giant planets \citep{pet18,Zink2023}, there is still a positive correlation that could lead to an enhancement in the frequency of small planets in Praesepe compared to the field (slightly sub-solar) \ktwo\ sample if not taken into account. Similarly, the field \ktwo\ sample includes a much larger fraction of higher mass stars than the Praesepe sample, and stellar mass has been shown to be anti-correlated with the frequency of small, short-period planets \cite[e.g.][]{mul15}. To account for these differences we include host star metallicity and stellar effective temperature terms in our parametric model.

For the parameter space probed by the nine Praesepe candidates (1--4~\rearth, 1--30 days), previous analyses using the older and much larger \emph{Kepler} sample have found that the population is well represented by a broken power law in orbital period with a break at $\sim$10 days \citep{mul15}, and a single power law in planet radius \cite[e.g.][]{pet18,muld18}. With only nine Praesepe planet candidates, our statistical power to simultaneously fit many parameters is low, and indeed an attempt to fit the full parametric model with a broken power law in orbital period and single power law planet radius was unable to meaningfully constrain the overall occurrence rate, let alone the exponents of the various power laws. However, by concentrating on hot sub-Neptunes, we can simplify the parametric fit in order to make more robust measurements. For the following analysis, we examine candidates with orbital periods 1--10~days, inwards of the expected period break and therefore able to be modeled with a single power law, and planet radii 1.7--4~\rearth, capturing the sub-Neptune population above the planet radius valley \citep{ful17,VanEylen2018,har20}. These planets are particularly susceptible to mass-loss mechanisms, due to their proximity to their host stars, and their weakly-bound gaseous upper atmospheres. In addition, the same parameter range is investigated in \cite{Zink2023}, which allows for a straightforward comparison to their results. 

In this space, we fit a parametric model $n$ of the form 

\begin{equation}
\label{eq:occ}
\frac{ d^4n}{d \log P \: d \log R \: d T_\textrm{eff} \; d \textrm{[Fe/H]} \;} = f\: \cdot P^{\beta}\: \cdot R_p^{\alpha}\: \cdot10^{\textrm{[Fe/H]}\cdot\lambda +  \frac{T_\textrm{eff}}{1000K} \cdot\gamma} 
\end{equation}

\noindent where $f$ captures the number of planets per star within the range of our sample, and $\alpha$, $\beta$, $\gamma$, and $\lambda$ are all free parameters, as in \cite{Zink2023}. We start by fitting the field \ktwo\ sample, allowing all tunable parameters to float. In our targeted stellar and planet parameter space, this leaves 149 planet candidates from \cite{Zink2021}. Examining the corner plot shows convergence of the MCMC parameter chains and highlights their independence. The fitted values are given in Table \ref{tab:coefficients}. For the field \ktwo\ sample, the occurrence rate is $16.54_{-0.98}^{+1.00}$\%, and we recover the strong negative radius scaling ($\alpha=-2.6\pm0.2$) and positive period power-law scaling ($\beta=2.1\pm0.1$) seen in previous studies (e.g. $\alpha=-2.7$ and $\beta=1.97$ from \cite{Zink2023}. We also see the expected weak but positive dependence on stellar metallicity ($0.19\pm0.09$), and the negative dependence on stellar effective temperature ($-0.072^{+0.018}_{-0.017}$)

\subsection{Praesepe}
\label{sec:praesepe}

In the Praesepe sample, we have three candidates in this parameter space (K2-100 b, K2-104 b, and K2-264 b), shown in the red dashed box in Figure \ref{fig:praesepe_candidates}. Even our simplified model is poorly constrained with only three candidates. We attempt to fit all five parameters ($f$, $\alpha$, $\beta$, $\gamma$, and $\lambda$), which gives an occurrence rate of 95\%, however, not unexpectedly, the fit does not converge, specifically for $\beta$ (the period power-law coefficient). This is perhaps driven by the fact that the three candidates occupy only the middle $\sim$half of the 1--10 day range. We proceed by first fixing the radius power law coefficient, $\alpha$, while fitting the remaining parameters, then fixing $\beta$ while fitting the remaining parameters, and finally fixing both $\alpha$ and $\beta$, using the values for $\alpha$ and $\beta$ from the full \emph{K2} field sample. Under the assumption that the radius and period power law coefficients are independent, which is likely appropriate over this limited region of parameter space, this allows us to explore the extent to which we can test for differences between these values for Praesepe compared to the full \emph{K2} data set, and in the overall occurrence rates.

The results for all fits are shown in Table \ref{tab:coefficients}. The resulting occurrence rates vary from $79^{+31}_{-25}$\% to $107^{+32}_{-26}$\% of GKM stars hosting a hot sub-Neptune \footnote{An occurrence rate of greater than 100\% indicates more than one hot sub-Neptune per star.}. The uncertainties are large, due to the small size of the sample, but the rates are still 2.5--3.5$\sigma$ higher than the 17\% occurrence rate recovered from the full \ktwo\ sample. In each case, the parameters describing the correlations with period, planet radius, stellar effective temperature and metallicity ($\alpha$, $\beta$, $\lambda$, and $\gamma$) lie within 1.3--1.6$\sigma$ of the values from the full \ktwo\ sample, indicating that our (small) Praesepe sample cannot demonstrate any significant differences in the strength of these correlations at younger ages. The only significant difference between the two samples is in the overall occurrence rate, which is significantly higher for the young small (1.7--4~\rearth), short-period (1--10 day) hot sub-Neptunes in Praesepe than for field stars. To test this, we perform a final fit where we fix all but the occurrence rate, $f$, assuming the null hypothesis that there is no difference in the correlations between the intermediate age and field stars, and find an occurrence rate of 100\%. 

\begin{figure}[ht!]
\begin{center}
\includegraphics[width=0.49\textwidth]{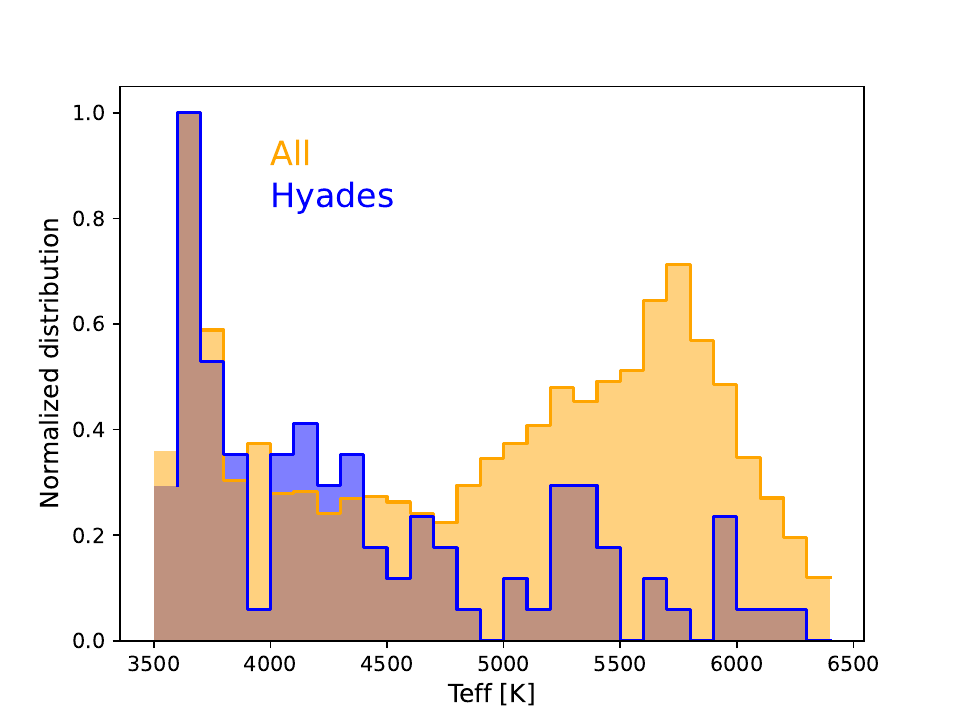}
\includegraphics[width=0.49\textwidth]{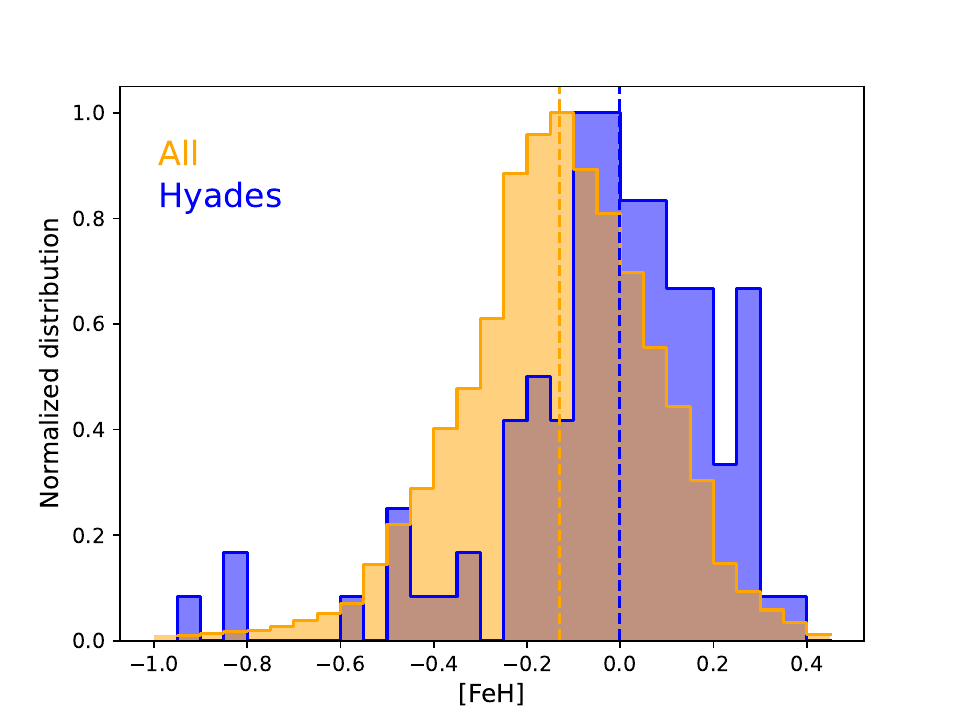}
\caption{As for Figure \ref{fig:praesepe_metallicity}, but for the significantly smaller Hyades GKM sample.\label{fig:hyades_distributions}}
\end{center}
\end{figure}

\subsection{Praesepe + Hyades}
\label{sec:hyades}

Given the high occurrence rate of hot sub-Neptunes found in Praesepe, we combine the Praesepe and Hyades targets to create a larger target list and test the result. The stellar effective temperature and metallicity distributions of the Hyades targets are shown in Figure \ref{fig:hyades_distributions}. The median metallicity of our 101 Hyades targets is [Fe/H] = -0.002, which is consistent with the cluster being slightly super-solar on average, given the typical uncertainty (0.235 dex) on our metallicity values and the small size of the sample. In the planet parameter space defined in Section \ref{sec:model}, we add one additional Hyades planet (K2-155 b) to our sample. We repeat the same sequence of fits as above, using the parametric model described in Section \ref{sec:model}. We first allow all five tunable parameters to float, which again does not converge for $\beta$ (the period power-law coefficient), the additional 6.3-day Hyades planet being close in period to the 5.8-day Praesepe planet and likely not providing additional constraints over the 1--10 day period range. We then fix the $\alpha$ parameter, the $\beta$ parameter, both parameters, and finally all four non-$f$ parameters. The fitted parameters $\alpha$, $\lambda$ and $\gamma$ lie within 0.4--1.8$\sigma$ of the values from the field \ktwo\ sample; the $\beta$ value is poorly constrained. As before, we cannot exclude the null hypothesis that the shape of the underlying planet population is the same for intermediate age stars as for field stars. The occurrence rates range from $90^{+31}_{-26}$\% to $121^{+30}_{-27}$\%, again indicating a significantly (2.8--3.8$\sigma$) higher rate of hot sub-Neptunes around these intermediate-age stars.

\begin{deluxetable*}{l|ccccccc}
\tablecaption{The measured parameters of the parametric model for the field \ktwo\ sample, the Praesepe sample, and the combined Praesepe$+$Hyades sample, over the two planet parameter ranges defined in the analysis. Bold values indicate parameters that were fixed in each analysis. $\dagger$: A repeat of the analysis on the line immediately above without the CDPP$<$1200~ppm cut, to confirm that it does not significantly affect the result.\label{tab:coefficients}}
\tablewidth{0pt}
\tablehead{
\colhead{Sample} &\colhead{$R_p$ [\rearth]} & \colhead{Period [d]} & \colhead{$f$} & \colhead{$\alpha$ [Radius]} & \colhead{$\beta$ [Period]} & \colhead{$\lambda$} & \colhead{$\gamma$}
}
\startdata
\ktwo\ & 1.7--4 & 1--10 & $0.1654^{+0.0100}_{-0.0098}$ & $-2.565^{+0.179}_{-0.209}$ & $2.103^{+0.109}_{-0.111}$ & $0.191^{+0.085}_{-0.085}$ & $-0.0718^{+0.0179}_{-0.0171}$\\
\cline{2-8}
  & 1.8--6 & 1--12.5 & $0.2076^{+0.0110}_{-0.0108}$ & $-3.697^{+0.144}_{-0.153}$ & $1.546^{+0.082}_{-0.082}$ & $0.200^{+0.084}_{-0.079}$ & $-0.0725^{+0.0163}_{-0.0167}$\\
\hline
Praesepe & 1.7--4     & 1--10 & $0.983^{+0.272}_{-0.253}$ & {\bf -2.565} & {\bf 2.103} & $1.06^{+0.71}_{-0.64}$ & $-0.246^{+0.176}_{-0.270}$\\
         & 1.7--4 & 1--10 & $0.7927^{+0.3081}_{-0.2482}$ & {\bf -2.565} & $2.301^{+1.430}_{-1.116}$ & $1.082^{+0.776}_{-0.708}$ & $-0.262^{+0.0193}_{-0.0285}$\\
          & 1.7--4     & 1--10 & $1.066^{+0.319}_{-0.260}$ & $-5.04^{+1.63}_{-1.91}$ & {\bf 2.103} & $1.16^{+0.68}_{-0.66}$ & $-0.277^{+0.185}_{-0.282}$\\   
          & 1.7--4     & 1--10 & $0.9448^{+0.4112}_{-0.3182}$ & $-5.470^{+1.752}_{-2.008}$ & $3.168^{+4.588}_{-1.727}{\dagger}$ & $1.333^{+0.729}_{-0.695}$ & $-0.301^{+0.193}_{-0.289}$\\
          & 1.7--4     & 1--10 & $0.9999^{+0.2911}_{-0.2455}$ & {\bf -2.565} & {\bf 2.103} & \bf{0.191} & \bf{0.0718}\\
          & 1.7--4     & 1--10 & $0.9474^{+0.2695}_{-0.2346}\dagger$ & {\bf -2.565} & {\bf 2.103} & \bf{0.191} & \bf{0.0718}\\
\cline{2-8}
          & 1.8--6 & 1--12.5 & $1.0451^{+0.2544}_{-0.2557}$ & {\bf -3.697} & {\bf 1.546} & $1.076^{+0.658}_{-0.668}$ & $-0.233^{+0.163}_{-0.276}$\\
          & 1.8--6 & 1--12.5 & $0.8736^{+0.3304}_{-0.2497}$ & $-6.242^{+1.585}_{-1.969}$ & $1.289^{+0.742}_{-0.630}$ & $1.159^{+0.707}_{-0.642}$ & $-0.265^{+0.175}_{-0.279}$\\
\hline
Prae$+$Hya & 1.7--4     & 1--10 & $1.1072^{+0.2601}_{-0.2771}$& {\bf -2.565} & {\bf 2.103} & $0.348^{+0.602}_{-0.535}$ & $-0.268^{+0.156}_{-0.222}$ \\
           & 1.7--4     & 1--10 & $1.2072^{+0.2977}_{-0.2678}$ & $-6.017^{+1.528}_{-1.772}$ & {\bf 2.103} & $0.455^{+0.583}_{-0.539}$ & $-0.295^{+0.164}_{-0.236}$ \\
          & 1.7--4     & 1--10 & $0.9094^{+0.3052}_{-0.2620}$& {\bf -2.565} & $5.226^{+5.174}_{-2.717}{\dagger}$ & $0.415^{+0.650}_{-0.600}$ & $-0.320^{+0.192}_{-0.236}$ \\ 
          & 1.7--4     & 1--10 & $0.9454^{+0.3645}_{-0.2934}$ & $-6.334^{+1.625}_{-1.962}$ & $4.582^{+3.877}_{-2.218}{\dagger}$ & $0.681^{+0.685}_{-0.584}$ & $-0.239^{+0.183}_{-0.239}$ \\
          & 1.7--4     & 1--10 & $1.1631^{+0.2578}_{-0.2475}$& {\bf -2.565} & {\bf 2.103} & \bf{0.191} & \bf{0.0718}\\
\cline{2-8}
           & 1.8--6     & 1--12.5 & $1.1640^{+0.2693}_{-0.2432}$ & {\bf -3.697} & {\bf 1.546} & $0.310^{+0.590}_{-0.514}$ & $-0.276^{+0.155}_{-0.226}$ \\
           & 1.8--6     & 1--12.5 & $0.8925^{+0.3054}_{-0.2547}$ & $-6.889^{+1.522}_{-1.953}$ & $2.070^{+1.094}_{-0.761}$ & $0.575^{+0.626}_{-0.599}$ & $-0.303^{+0.172}_{-0.238}$ \\           
\enddata
\end{deluxetable*}

\section{Discussion}
\label{sec:disc}

We investigated the occurrence rate of hot sub-Neptunes in the intermediate-age Praesepe and Hyades clusters observed by \ktwo. We derived significantly higher occurrence rates (79-121\%) for the small (1.7--4~\rearth), short-period (1--10 day) planets in our Praesepe and combined Praesepe$+$Hyades datasets than for our full field \ktwo\ sample (17\%). Although less work has been done to age-date \ktwo\ field stars than \emph{Kepler} stars, examining the ages of the field \ktwo\ planet host stars reveals of median of 5.6~Gyr, and a 68 percentile range of 2.8--9.0~Gyr \citep{PSCP}, which is similar to the range of ages found for \emph{Kepler} targets \cite[see, e.g.][]{David2022}. We therefore identify a significant correlation between stellar age and the occurrence rate of hot sub-Neptunes.

\begin{figure}[ht!]
\begin{center}
\includegraphics[width=\textwidth]{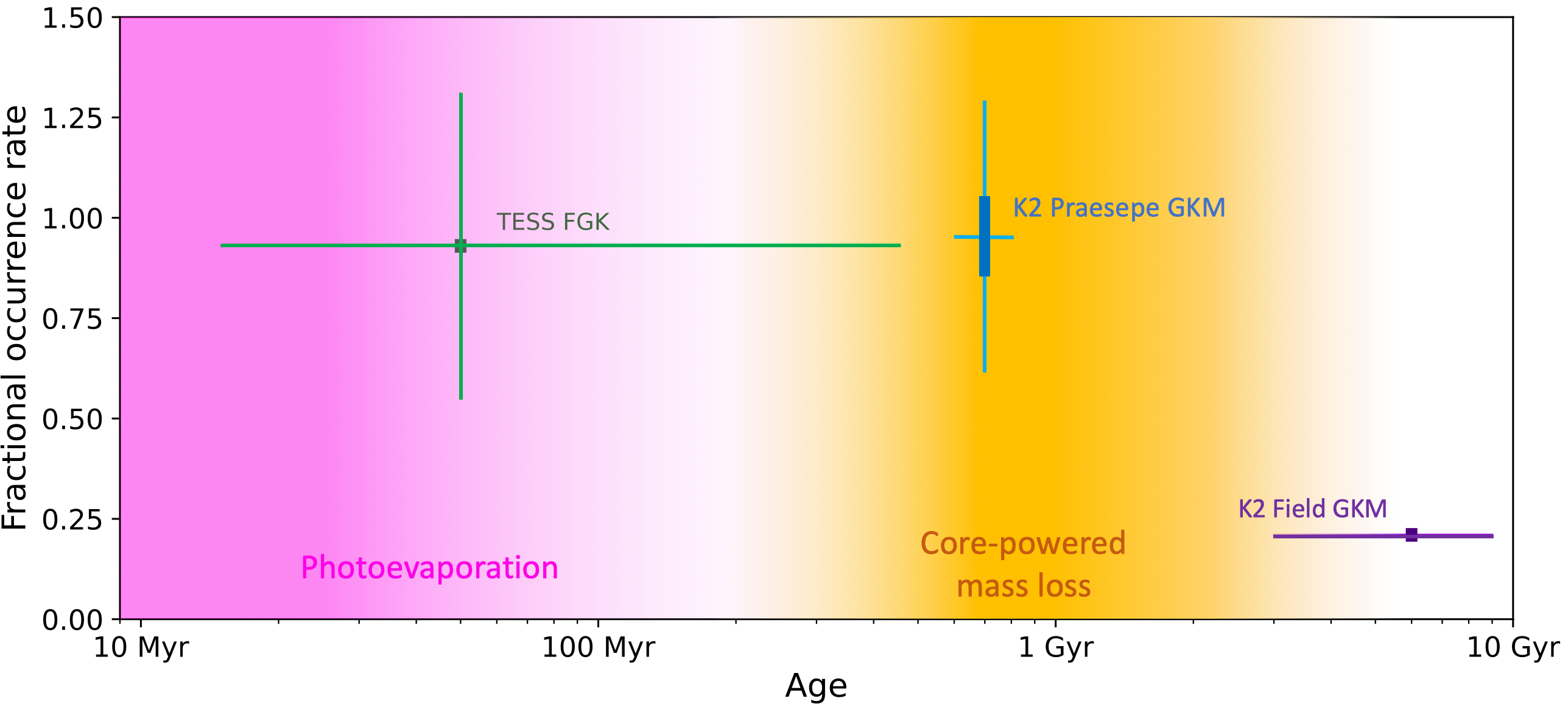}
\caption{The occurrence rates of hot sub-Neptunes with time, from the young TESS FGK stars (15--450~Myr) from \cite{Fernandes2023}, to the intermediate age (600--800~Myr) \ktwo\ Praesepe GKM sample studied here, to the field \ktwo\ GKM stars (3--9~Gyr). The age uncertainties capture the width of the age distributions in each case. The shaded regions indicate the putative timescales for the photoevaporation ($<$100~Myr) and core-powered mass loss processes (0.5--2~Gyr). The persistent high rate of hot sub-Neptunes at $\sim$700~Myr is more consistent with predictions of core-powered mass loss sculpting this population than with photoevaporation.
\label{fig:occrate_with_time}}
\end{center}
\end{figure}

Recently, \cite{Fernandes2023} also found a hint of a higher occurrence rate for small, short-period planets around young stars. They analyzed 5 nearby young clusters observed by \emph{TESS}, with ages spanning from 15--450~Myr, and a median age of $\sim$45--50~Myr. Over a slightly larger planet parameter space (planet radii from 1.8--6~\rearth\ and orbital periods from 1--12.5~days), and a slightly higher mass sample of stars than the GKM sample studied here (FGK stars from 0.55--1.63~\msun), they find an occurrence rate of $93\pm38$\%\footnote{Re-analysis wih a GKM sample produced an almost identical number of $96\pm39$\% (private communication).}, with the caution that they began by analyzing the clusters in which planets had already been discovered, which may have biased their derived occurrence rate. We repeat our analysis described above in this larger parameter space, as shown in Table \ref{tab:coefficients} (albeit for our slightly different stellar mass range). This has the effect of removing the smallest planet from our previous sample (K2-104b; 1.77~\rearth), and adding a longer period Praesepe planet \cite[K2-95 b, 10.13~days;][]{Obermeier2016}. First, we fit the full \ktwo\ sample, finding an occurrence rate slightly increased from 17\% to 21\%. With the addition of the longer period planet, we are able to fit all five tunable parameters, and do so for the Praesepe and Praesepe$+$Hyades samples, finding occurrence rates of $87^{+33}_{-25}$\% and $89^{+31}_{-25}$\%, respectively. 
These are 2.7$\sigma$ higher than the full \ktwo\ occurrence rate, and are in excellent agreement with the value from the younger \emph{TESS} sample, shown in Figure \ref{fig:occrate_with_time}, lending credence to the result that the occurrence rate of hot sub-Neptune planets is significantly higher around younger stars ($<$1000~Myr) than around 3--9~Gyr-old field stars. For consistency with our previous analysis, we also perform fits where the $\alpha$ and $\beta$ parameters are fixed to the values found for the full \ktwo\ sample, shown in Table \ref{tab:coefficients}. For both the Praesepe and Praesepe$+$Hyades samples this has the effect of increasing the overall occurrence rate and increasing the strength of the negative trend with radius.

Identifying a correlation between stellar age and planet occurrence rates could imply one of several things. Potentially, there is some ongoing process after the birth of a planetary system, independent of when that system is born, that acts to sculpt and evolve planets over time, such that the planetary systems we observe in young clusters evolve to become the planetary systems we see in the field \ktwo\ sample. Or, that stars born at different ages of our Galaxy are born in sufficiently different conditions as to create different types and architectures of planetary systems in the first place. Here, we discuss these possibilities.

\subsection{Planet atmosphere evolution}

One intriguing property of small, short-period planets is their bimodal distribution in planet radius \citep{ful17,VanEylen2018,har20}, or perhaps more accurately trimodal in planet density \citep{Luque2022}. Many of the ideas proposed to explain this distribution involve mass loss from planets early in their histories, including through photoevaporation \citep{Lopez2012,Lopez13,Owen2013,Owen2017,Rogers2021} or core-powered mass-loss \citep{Ginzburg2018,gup19,Gupta2020,Gupta2021}. Perhaps the very high rate of hot sub-Neptunes found in Praesepe and Hyades is in the process of reducing to the smaller rate we see in the \ktwo\ field stars.


In the photoevaporation scenario, the gaseous atmosphere is stripped away by high energy (XUV) radiation from the host star. The extent of the mass loss, and the timescale over which it occurs, are driven by the initial planetary core mass, the initial envelope mass fraction, the orbital separation of the planet, and the stellar spectral energy distribution. \cite{Lopez2012} investigated the role of photoevaporation mass-loss for a specific system, finding that the 1.8~\rearth\ planet Kepler-11~b could have lost a factor of 2--3 in radius and 1.5--5 in mass between $\sim$100~Myr and 10~Gyr. \cite{Lopez13} extended this study, predicting that planets in the range 1.8--4.0~\rearth\ should become significantly less common on orbits $<$10 days. Subsequently, \cite{Rogers2021} concluded that although some super-Earths are born rocky, approximately four times as many are born with large H/He atmospheres that are subsequently stripped by photoevaporation, a number that is tantalizingly in agreement with our finding of 4--5 times as many young sub-Neptunes as old sub-Neptunes, if a significant fraction of the deficit evolve to become super-Earths. 

In the core-powered mass loss scenario, the gaseous atmosphere can be blown away by both the planet's primordial energy from formation and bolometric heating from the star. If the formation energy is similar to or greater than the gravitational binding energy of the atmosphere, the cooling luminosity of the planet may blow off a H/He atmosphere entirely. For planets with masses between Earth and Neptune, the core temperatures can be 10,000--100,000~K, which can take up to several Gyr to cool \citep{Ginzburg2016}. \cite{Gupta2020} found that the average size of sub-Neptunes decreases significantly with age, and commensurately that the relative occurrence rate of sub-Neptunes decreased by a factor of 2--2.5 over the timescale of the mass loss (see their Fig.~10), which is consistent at the 1--2$\sigma$ level with our results.

One obvious question is whether our finding, that the occurrence rate of hot sub-Neptunes is still very high at $\sim$700~Myr, before dropping by a factor of up to 5 at several Gyr, agrees with the putative timescales of these mass-loss mechanisms. \cite{Owen2013,Owen2017} investigated the timescales of photoevaporation, finding that most atmospheric erosion happens within the first 100~Myr \cite[see, e.g., Fig. 2 of][]{Owen2017}. 
This would imply that planets shrink in size somewhat sooner than our Praesepe$+$Hyades result, but is not at odds with the younger TESS result. Core-powered mass loss is expected to occur over much longer timescales than photoevaporation, typically 0.5--2~Gyr \citep{Ginzburg2016,gup19,Gupta2020}, which is much more consistent with our finding that at 0.7~Gyr these planets are still large. \cite{Gupta2020} state \emph{``we expect a drastic change in the planet size distribution between stars that are younger and older than the typical core-powered mass-loss time-scale, which is of the order of a Gyr. Due to these long mass-loss time-scales, we predict the transformation of sub-Neptunes into super-Earths to continue over Gyr time-scales''}. Indeed, early examination of individual planets being detected in the Praesepe and Hyades \ktwo\ observations noted that they appeared to be larger, on average, than planets orbiting older stars of the same mass observed with \emph{Kepler} \cite[see, e.g. Fig. 12 of][]{Rizzuto2018}, which could have been evidence for ongoing thermal contraction of their atmospheres, and/or could have hinted at a preference for core-powered mass loss \citep{Gupta2020}. However, without understanding the completeness of the survey detecting those planets, this could just as easily have been a selection bias, where the noisier \ktwo\ light curves limited the detection to larger planets when compared with the more well-behaved \emph{Kepler} light curves. \cite{Berger2020} and \cite{Sandoval2021} used isochronal ages to divide the \emph{Kepler} planet sample into coarse age bins, and found that the relative frequency of sub-Neptunes to super-Earths decreases with increasing age, i.e. that either the number of sub-Neptunes was decreasing, the number of super-Earths was increasing, or both. Here we extend the \emph{Kepler} ages to the young clusters observed by \ktwo\ and find that there is indeed a ``drastic change'' (a factor of $\sim$4--5) in the absolute frequency of sub-Neptunes between stars that are younger and older than $\sim$1~Gyr. This result, specific to the Praesepe and Hyades clusters studied here, is more consistent with the published predictions of the core-powered mass loss origin for the planet radius valley rather than a photoevaporation origin, or perhaps points to some missing process in the photoevaporation model that would act to slow down the expected mass loss rate. Indeed, recent work by \cite{Owen2023} conversely theorizes that planets that have undergone core-powered mass loss can then transition to photoevaporation, depending on the initial envelope mass fraction; whichever the order, the bulk of the mass loss is indicated to happen after the first 0.5~Gyrs.

Another factor that must be considered is the natal disk opacity, which plays an important role in both photoevaporation and core-powered mass loss processes. Planets that have formed from material that is richer in metals have the ability to maintain their envelopes for a longer period of time due to their increased heat capacity, enabling efficient cooling of their outer envelopes. Given that our cluster planet sample is more metal-rich than the [Fe/H] $\sim-0.2$ of \emph{Kepler} \citep{don14} and the [Fe/H] $\sim0.0$ of the field \ktwo\ sample, it is important to consider how this may affect our conclusions. According to \citet{owen18}, the envelope mass loss rate ($\dot{M}$) dictated by the photoevaporation model scales with the disk metallicity ($Z$) as $\dot{M}\propto Z^{-0.77}$, while the mass loss originating from the residual heat in the planet's core scales as $\dot{M}\propto Z^{-1}$ \citep{Gupta2020}. This implies that the removal of sub-Neptune atmospheres will be delayed in a metal-rich environment, increasing the expected timescales of photoevaporation and core-power mass loss processes by $1.4\times$ and $1.6\times$, respectively.\footnote{The increased timescales displayed represent upper limits on the respective timescale corrections. The overall mass loss efficiency is also likely impacted by this change in composition, truncating the overall magnitude of the mass loss processes.} However, the surplus of hot sub-Neptunes at 600--800~Myr is more than $6\times$ the fiducial timescale for photoevaporative mass loss ($\sim100$ Myr). Therefore this potential delay, arising from the cluster's metallicity enhancement, cannot account for the abundance of hot sub-Neptunes identified in this study.

Finally, our sample covers a wide range of stellar masses, from 0.3 to 1.3 solar masses. The evolutionary timescales discussed above have been derived for an FGK sample of stars. The inclusion of low mass stars in our sample may modify the time frame over which hot sub-Neptunes are converted to super-Earths. Under the action of the photoevaporation framework, small M dwarfs experience a heightened integrated XUV flux, accelerating and/or increasing the magnitude of the mass loss process \citep{Rogers2021b}. Qualitatively, this would reduce the expected timescale needed to dissociate the planetary envelopes of sub-Neptunes, culminating in an overall occurrence reduction in the 600--800~Myr clusters. However, our results show that the opposite is true. In contrast, the core-powered mass loss evolutionary process relies on the star's total luminosity and therefore expects decelerated and/or reduced mass loss around less massive stars \citep{Gupta2020}. The inclusion of M dwarfs should increase the expected mass loss timescale and leave a larger reservoir of short-period sub-Neptunes at 600--800~Myr, matching the observation of this survey. Overall, the complexities of stellar mass and composition unique to our sample are directionally consistent with a core-powered mass loss mechanism, providing further evidence in favor of this formation pathway.  

\subsection{Primordial planet formation}

An alternative explanation for the overabundance of hot sub-Neptunes in Praesepe and Hyades compared to the field \ktwo\ sample is that there could be systematic evolution of planetary system formation and architectures with different environmental/Galactic conditions over cosmic time. There are a number of known correlations between small, short-period planet occurrence rates and their host star properties---including properties that have evolved over the lifetime of the Galaxy, as the natal conditions for each generation of stars is impacted by previous generations. One such property is the host star metallicity---stars that are born today are typically formed in giant molecular clouds that have been enriched by the supernova remnants of earlier stars, in a more heavy-element-rich environment than in early Galactic times. This is reflected in the slightly super-solar average metallicities of Praesepe and Hyades, but was also explicitly fit as part of our parametric model, so presumably what we are seeing here is an additional effect. Both observations and models of galaxy formation in Milky Way-like galaxies suggest that stars born at earlier times are typically born in more strongly dynamically clustered environments \citep{Lee2020,Grudic2023}, are much more strongly irradiated externally by elevated galactic star formation rates \citep{Tacconi2020,guszejnov:fire.gmc.props.vs.z}, with likely larger ionizing cosmic ray fluxes \citep{Hopkins2021} and systematically warmer natal gas clouds \citep{guszejnov:fire.gmc.props.vs.z}. There is a long history of work on how this may or may not influence the low-mass end of the stellar initial mass function, but until recently little work on its possible effects on planet formation. 
Possibly, early and frequent interactions with nearby stars destabilize planetary systems to the extent that the overall occurrence rate, even for planets on the close-in orbits probed by our analysis, is reduced. \cite{Zink2023} recently discovered a strong negative correlation between the occurrence rate of 1.7--4~\rearth, 1--10 day planets (the same planet parameter space probed here) and the maximum height above the Galactic plane that the host stars reach as they oscillate through the plane. This height can be used as a proxy for thick disk (older, more metal poor, more likely to have formed in more clustered stellar environments) versus thin disk (younger, more metal rich, less likely to have formed in more clustered environment) membership, and since \cite{Zink2023} also accounted for metallicity, points to an additional line of evidence that older stars have fewer small, short-period planets when compared to younger stars, although as here there was no systematic difference identified in the shape of the planet population. Although there has been less work on quantitative predictions for these primordial planet formation models compared to the atmosphere evolution models, nothing in this work disfavors such explanations as a category.







\section{Conclusions}

We analyze \ktwo\ observations of the Praesepe and Hyades clusters, and measure a significantly higher occurrence rate for hot sub-Neptunes orbiting intermediate (600--800~Myr) GKM stars than for the older (3--9~Gyr) GKM field stars. Notably, our result persists after the inclusion of additional data, when varying the boundaries of the planet parameter space under inspection, and regardless of whether the shape of the population is assumed to be the same as that of the field stars or allowed to vary. If the decline in the number of hot sub-Neptunes with age is a result of planet evolution, the high rate of these planets at intermediate ages is more consistent with the predicted timescales of the core-powered mass loss mechanism than the photoevaporation mechanism. If it is instead a reflection of the primordial planet populations in these clusters compared to the field stars, it may point to some other trend with planet formation as a function of Galactic age, dynamical or radiation history, or cluster environment. One interesting test would be to search Praesepe and Hyades for longer period planets, such as with radial velocity, astrometry, or direct imaging. The putative mechanisms for planet evolution happen predominately at close-in distances to the host star---if the outer regions of the intermediate age planetary systems look similar to the older planetary systems, that could point to evolution being the explanation, rather than primordial differences in planetary system architectures.

\begin{acknowledgments}
This research has made use of the NASA Exoplanet Archive, which is operated by the California Institute of Technology, under contract with the National Aeronautics and Space Administration under the Exoplanet Exploration Program.This paper includes data collected by the Kepler mission and obtained from the MAST data archive at the Space Telescope Science Institute (STScI). Funding for the Kepler mission is provided by the NASA Science Mission Directorate. STScI is operated by the Association of Universities for Research in Astronomy, Inc., under NASA contract NAS 5–26555. This material is based upon work supported by the National Aeronautics and Space Administration under Agreement No. 80NSSC21K0593 for the program “Alien Earths”. The results reported herein benefited from collaborations and/or information exchange within NASA’s Nexus for Exoplanet System Science (NExSS) research coordination network sponsored by NASA’s Science Mission Directorate. JZ acknowledges the support provided by NASA through the Hubble Fellowship grant HST-HF2-51497.001 awarded by the Space Telescope Science Institute, which is operated by the Association of Universities for Research in Astronomy, In., for NASA, under the contract NAS 5-26555. Support for PFH was provided by NSF Research Grants 1911233 \& 20009234, NSF CAREER grant 1455342, NASA grants 80NSSC18K0562, HST-AR-15800.001-A.
\end{acknowledgments}

%

\vspace{5mm}
\facilities{Kepler}


\software{astropy \citep{2013A&A...558A..33A,2018AJ....156..123A},  
          ExoMult \citep{zin19}, BANYAN $\Sigma$ \citep{Gagne2018}
        }






\bibliography{sample631}{}
\bibliographystyle{aasjournal}



\end{document}